\documentclass[pra,twocolumn,superscriptaddress]{revtex4-1}
\usepackage{graphicx}
\usepackage{subfigure}
\usepackage{amsmath}
\usepackage{mathrsfs}
\usepackage{amsthm}
\usepackage{amssymb}
\usepackage{hyperref}
\usepackage{xcolor}
\usepackage[OT4]{fontenc}

\usepackage{ulem}
\usepackage{dcolumn}
\usepackage{bm}

\hypersetup{colorlinks=true, linkcolor=black, citecolor=black, urlcolor=blue}
\allowdisplaybreaks

\begin{document}

\author{Pawe{\l} Zin}
\affiliation{National Centre for Nuclear Research, ul. Pasteura 7, PL-02-093 Warsaw, Poland}

\author{Maciej Pylak}
\affiliation{National Centre for Nuclear Research, ul. Pasteura 7, PL-02-093 Warsaw, Poland}
\affiliation{Institute of Physics, Polish Academy of Sciences, Aleja Lotnik\'ow 32/46, PL-02-668 Warsaw, Poland}

\author{Mariusz Gajda}
\affiliation{Institute of Physics, Polish Academy of Sciences, Aleja Lotnik\'ow 32/46, PL-02-668 Warsaw, Poland}
\email{gajda@ifpan.edu.pl}

\title{Revisiting a stability problem of two-component quantum droplets}

\newcommand{\x}{{\bf r}}

\begin{abstract}
We study the problem of the stability of a two-component droplet. The standard solution Ref. \cite{Petrov15} is based on a particular form of the mean field energy functional, in particular on the assumption of vanishing large energy hard mode contribution. This imposes a constraint on the densities of the two components. The problem is reduced to stability analysis of a one component system. As opposed to this, we present a two component approach including possible hard mode excitations. We minimize the energy under conditions corresponding to the experimentally relevant situation where volume is free and atoms can evaporate. For the specific case of a two component Bose-Bose droplet we find approximate analytic solutions and compare them to the standard result. We show that the densities of a stable droplet are limited to a range depending on interaction strength, in contrast to the original unique solution.
\end{abstract}

\maketitle

\section{Introduction}
Quantum droplets, as predicted in the seminal paper by D.~Petrov \cite{Petrov15}, are self bound systems of a mixture of two Bose-Einstein condensates under such conditions that interspecies attraction drives them towards collapse. The stabilizing agent is the Lee-Huang-Young (LHY) energy \cite{LHY} originating from quantum fluctuations of the Bogoliubov vacuum. The stability analysis presented in \cite{Petrov15} is based on the observation that a stable droplet can be formed if interactions are chosen in such a way that the mean field energy almost vanishes. In fact the system should be effectively very weakly attractive, so that the instability is suppressed by a small contribution of quantum fluctuations -- the LHY term. 

Soon experiments came. The first experiments with mixtures of Potassium atoms in two different internal states  \cite{Cabrera18,Fattori18,Tarruell18} as well as the recent achievement of quantum droplets in heteronuclear bosonic mixtures \cite{Modugno19a} well agree with the theoretical predictions. Moreover, a similar stabilization mechanism, originating from quantum fluctuations \cite{Pelster}, occurred to be responsible for stabilization of elongated dipolar condensates of Dysprosium \cite{Pfau16,Pfau16a,Kadau16,Ferrier-Barbut16} and Erbium \cite{Chomaz16} atoms.  For a review on the present state of quantum droplet physics see \cite{malomed1}.

Standard stability analysis of a two component Bose-Bose mixture presented in Ref. \cite{Petrov15} is based on the observation that the mean field energy density $\varepsilon(n_1,n_2)$ is a quadratic form of densities of both components $n_1$, $n_2$, and can be brought to the diagonal form $\varepsilon(n_1,n_2)=\lambda_{+}n_+^2+\lambda_- n_-^2$. Explicit form of $n_\pm$ and $\lambda_\pm$ is not needed here, it can be found in \cite{Petrov15}. $n_+$ and $n_-$ are densities of hard and soft modes respectively, both being linear combinations of $n_1$ and $n_2$. The dominant contribution to the energy density comes from the hard mode, which owes its name to the hard mode interaction strength $\lambda_+$ which significantly dominates the soft mode strength, $\lambda_-$, i.e.  $\lambda_{+} \gg |\lambda_{-}|$. The soft mode is unstable $\lambda_{-}<0$ but the instability is weak and can be tamed by the small LHY energy. This energy is negligible in typical experimental arrangements but becomes crucial when the mean field energy nearly vanishes. This can happen if the hard mode contribution is very small. In \cite{Petrov15} the approximation $n_+(n_1,n_2) = 0$ is used. The densities of both species become dependent then.

This way the problem of a mixture can be reduced to an effectively single component case -- the energy density becomes a functional of a single function only. To get the equilibrium densities of a self bound droplet it is enough to notice that in a free uniform system the pressure vanishes. This condition determines unambiguously the density of the droplet. Although dilute ($\sim 10^{14} - 10^{15} {\rm cm}^{-3}$), quantum droplets behave like liquids. Their densities are fixed by interaction and in the limit of an infinite system they do not depend on atom number. 

In this paper we want to address the issue of stability of a two component droplet going beyond approximations which rely on the distinction between hard and soft modes and neglect the former one. The two component approach shows the standard result from a broader perspective. Moreover, it allows to find a contribution of hard mode excitations to the ground state energy of the stable droplet. We want to add that the hard mode contribution to the energy may be important not only in the ground state. It reveals itself especially in collisions of droplets \cite{pylak_tobe}.  

We address the issue of the stability of the mixture in a quite general setting, however, we have in mind a two component system of ultracold bosonic atoms. The stability conditions are formulated in a form not assuming any particular energy density functional. These conditions are determined by constraints imposed by the typical physical situation. The question of a global unconstrained minimum has a simple but trivial answer if the analysis is limited to the mean field approach -- (i) in an effectively attractive case the system collapses and both densities become infinite, (ii) or on the other hand if the system is a repulsive one, the atoms expand to infinity and their densities vanish. The collapse predicted on the mean field level in fact signifies that the description used does not account for physical processes in this situation. Formation of bound molecules, larger complexes or solidification is expected then.

We study a typical experimental situation where initially $N_1$ atoms of the first component are mixed with $N_2$ atoms of the second component in an external trap. After tuning the interactions to the region in which a stable droplet is expected, the external potential is removed. Eventually a droplet is formed. This is a scenario which defines plausible physical constraints. Our goal is to find the densities of a stable system formed this way and/or the final number of atoms in each component. Note that the final number of atoms need not be necessarily the same as the number of atoms mixed initially, as some may evaporate. 

\section{Constraints}

In this section we formulate constraints defining a stable self bound system taking into account the typical physical situation. Although we explicitly refer to a Bose-Bose mixture we want to stress that the approach presented here can be, after minor adjustments, applied to other mixtures provided that their potential energy depends on densities of both species only. In particular we have in mind Bose-Fermi mixtures as studied in \cite{Rakshit19,Rakshit19low}. 

The interaction energy density of a mixture of Bose-Einstein condensates, for fixed interaction strength, is a function of densities of the two components $n_1, n_2$, $\varepsilon =\varepsilon(n_1,n_2)$. The densities are related to corresponding wavefunctions $n_i = |\psi_i|^2$ which are normalized to the number of particles $\int d{\x} |\psi_i({\x})|^2 = N_i$. The total energy density with kinetic energy included is:
\begin{equation}
\label{e_total}
{\cal E}=-\frac{\hbar^2}{2m}\sum_{i=1}^2 \nabla \psi_i^* \nabla \psi_i +\varepsilon(n_1,n_2),
\end{equation} 
where we assumed equal masses $m$ of the two kind of atoms involved. The choice of equal masses has one serious advantage --  it allows for (to a large extent) analytical treatment. The entire procedure is also valid for different masses of both species, however, numerics is needed then. 

For a droplet to be formed the energy functional $\varepsilon$ should contain a positive contribution from repulsive intraspecies interactions as well as a negative contribution from attractive interaspecies interactions. 

A corresponding time-dependent set of two coupled Gross-Pitaevskii (GP) equations describing the dynamics of both components can be easily obtained by minimizing the action ${\cal S} = \int d {\x} \int dt  {\cal L}$, where the Lagrangian density is ${\cal L}  = \hbar {\cal R}e(i  \sum_j  \psi_j^* \partial_t \psi_j) - {\cal E}$:
\begin{equation}
i \hbar  \frac{\partial}{\partial t} \psi_i (\x) = \left[ -\frac{\hbar^2}{2m}\Delta + \frac{\delta \varepsilon(n_1,n_2)}{\delta n_i} \right] \psi_i(\x)
\label{e-gp1}
\end{equation}
By the standard substitution $\psi_i(\x,t)=e^{-i \mu_i t/\hbar}\psi_i(\x)$ the time dependent equations lead to a set of two coupled stationary GP equations:
\begin{equation}
\left[ -\frac{\hbar^2}{2m}\Delta + \frac{\delta \varepsilon(n_1,n_2)}{\delta n_i} \right] \psi_i(\x)  =  \mu_i \psi_i (\x)
\label{e-gp2}
\end{equation}
where the eigenvalues $\mu_i$ are chemical potentials.

We assume that interactions are tuned in such a way that the system is effectively very weakly attractive and is on the collapse side of the stability diagram \cite{Rzazewski13,Oldziejewski16}.
As shown by D. Petrov \cite{Petrov15}, if the energy of quantum fluctuations is included in $\varepsilon(n_1,n_2)$ in addition to the aforementioned mean-field interaction energy, the collapse may be avoided and a liquid droplet of volume $V$ can be formed. This however can only happen if the interactions are appropriately tuned. Moreover, the numbers of available atoms in every component must be in a right proportion. 



In general the number of particles forming a droplet is different from the number of atoms $N^{ini}_1, N^{ini}_2$ prepared initially in the trap and used in the formation process. After the trapping potential is removed the system is free. There are no external mechanisms of controlling the volume or number of atoms in the system. The excessive particles will be ejected and will not contribute to the total energy. We do not assume interaction of the system with any external reservoir of particles, the number of particles forming a droplet may not grow larger than the initial $N^{ini}_1$ and $N^{ini}_2$. These are the only physical constraints we impose on the system.

The question we want to answer here is: {\it  which stable system (for fixed interactions) can be formed having at disposal $N^{ini}_1$ atoms of the first kind and $N^{ini}_2$ atoms of the second kind?}  

Stable solutions of GP equations Eq.(\ref{e-gp1}) should correspond to a minimum of energy. If it is a global minimum the system is absolutely stable. Metastable states correspond to a local minimum of energy. A potential barrier separates the system from the global minimum. The total energy of the system is:
\begin{eqnarray}
E(N_1,N_2) = \int d{\x} \, {\cal E}(\x).
\end{eqnarray}
Chemical potentials $\mu_i$ appear in Eqs.(\ref{e-gp1}, \ref{e-gp2}) as eigenenergies of stationary solutions of the GP equations. It is a simple exercise to verify that these eigenenergies $\mu_i$, as it should be in the case of true chemical potentials, describe a response of the total energy of the system to a change of particle number:
\begin{equation}\label{mui1}
\frac{\partial E}{\partial N_i}=\mu_i
\end{equation}
Because the energy density functional, Eq.(\ref{e_total}), accounts for the kinetic energy, the density of a droplet decays exponentially at its surface. Density profiles follow directly from solutions of the stationary GP equations Eq.(\ref{e-gp2}) and volume of the droplet is not an additional parameter.

If the system is stable, i.e. if its energy $E(N_1,N_2)$  corresponds to some minimum, then infinitesimally small change of atom number in any component must increase its energy:
\begin{equation}
\label{stab1}
dE=\frac{\partial E}{\partial N_1} dN_1+\frac{\partial E}{\partial N_2} dN_2 = \mu_1 dN_1 + \mu_2 dN_2 >0
\end{equation}
In a typical experimental situation the number of atoms may only decrease, i.e. $dN_i <0$. If any of the chemical potentials were positive the system would decrease its energy by evaporating some particles of the corresponding kind. Therefore the constraints we impose on a stable droplet are:
\begin{eqnarray}
\label{stab_cond1}
\mu_1 & < & 0, \\
\label{stab_cond2}
\mu_2 & < & 0.
\end{eqnarray}
If in a given state both chemical potentials are negative then there is no state of lower energy in its close neighbourhood. We are going to exploit these conditions in the following. 

Note, however, that in the above we analyzed only stationary states i.e. states being the solution of stationary GP equation (\ref{e-gp2}). We did not consider dynamical stability of the solution against some small perturbations. It is known that there exists stationary localized droplet solutions which are however dynamically unstable - a small initial perturbation grows exponentially in time \cite{malomed2}. In what follows we do not consider the issue of dynamical stability.

\section{Bose-Bose droplets}
\subsection{Region of stability}
In the general approach sketched above a kinetic energy was included. This way we accounted for surface tension providing a necessary pressure to stabilize the system. Unfortunately including kinetic energy leads to differential equations which cannot be treated analytically in more detail in the general case.

To get some better insight into the problem of stability of a droplet we simplify our analysis and assume that the system is large and the surface energy is much smaller than the interaction energy so that it can be neglected. This approximation is known as the Thomas-Fermi approximation. It amounts to assuming that ${\cal E} = \varepsilon(n_1,n_2)$. Such a system is uniform, has well defined volume $V$ and constant densities $n_i=N_i/V$. The energy density of a mixture of two quantum-degenerate Bose gases is of the form: 

\begin{eqnarray}
\varepsilon(n_1,n_2) /\left(\frac{4\pi \hbar^2}{m}\right) = \frac{1}{2}\sum_{i=1}^2{a_{ii}n_i^2}-a_{12}n_1n_2+\nonumber\\
+c (a_{11} n_1 + a_{22} n_2)^{5/2}
\end{eqnarray}
where $a_{ij}>0$ are scattering lengths. The first term describes repulsive interspecies interactions while the second one corresponds to interspecies attraction. The last term is the LHY energy contribution, $c=64/(15\sqrt{\pi})$ By introducing the parameter $\delta a = -(a_{12}+\sqrt{a_{11}a_{22}})>0$ the energy density can be brought to the form
\begin{eqnarray}
\varepsilon(n_1,n_2) /\left(\frac{4\pi \hbar^2}{m}\right) = \frac{1}{2} (\sqrt{a_{11}} n_1 - \sqrt{a_{22}} n_2 )^2\nonumber\\
-\delta a n_1 n_2 + c (a_{11} n_1 + a_{22} n_2)^{5/2}
\label{e-density}
\end{eqnarray}
where "hard mode" and "soft mode" are clearly visible.  We assume that $\delta a \ll a_{11}, a_{22}$, i.e. that the collapse instability is weak and a small LHY term can balance it. This assumption ensures that the system is weakly interacting. 

The total energy of the uniform system is $E_u(N_1,N_2,V)=V\cdot\varepsilon(N_1/V,N_2/V)$. It depends not only on the number of particles $N_i$ but also on the volume $V$. Differential change of energy due to infinitesimal change of volume and particle number is:
\begin{equation}
\label{dE}
d E_u(N_1,N_2,V) = -p dV + \mu_{1,u} dN_1 +  \mu_{2,u} dN_2
\end{equation}
where 
\begin{equation}
p=-\frac{\partial E_u}{\partial V},
\end{equation} 
is a pressure, while 
\begin{equation}
\label{mui2}
\mu_{i,u} = \frac{\partial E_u}{\partial N_i} = \frac{ \partial \varepsilon}{\partial n_i},
\end{equation}
are chemical potentials of the species. Note that we used different symbols from those in Eq.(\ref{mui1}).

For a uniform free system, as opposed to a system with a surface, we get an additional constraint: a droplet will stabilize its volume if internal pressure vanishes: 
\begin{equation} 
\label{pressure}
p=-\frac{\partial E_u}{\partial V}= \mu_{1,u} n_1 +\mu_{2,u} n_2-\varepsilon(n_1,n_2)=0
\end{equation}
Equation~(\ref{pressure}) allows to find the volume of a droplet as a function of particle number $V=V_u(N_1,N_2)$. 
\begin{equation}
V_u(N_1,N_2)^{1/2} = \frac{ 3 c (a_{11} N_1 + a_{22} N_2)^{5/2} }{ 2 \delta a N_1 N_2 - 
(\sqrt{a_{11}} N_1 - \sqrt{a_{22}} N_2 )^2 }.   
\label{volume}  
\end{equation}
Physical solutions (i.e. solutions with real and positive $V_u(N_1,N_2)$) of Eq.~(\ref{pressure}) exist if: 
\begin{equation}
|\sqrt{a_{11}} N_1 - \sqrt{a_{22}} N_2 |< \sqrt{2 \delta a N_1 N_2}
\label{volume+}
\end{equation}
The first important observation is that the right hand of inequality  Eq.(\ref{volume+}) significantly reduces the possible variation of the ratio $N_1/N_2$, because $\delta a \ll a_{11}, a_{22}$. Thus a term 
$\sqrt{a_{11}N_1} - \sqrt{a_{22}N_2}$ must be very small. To quantify this difference we introduce a small parameter $\delta b = \frac{\delta a}{\sqrt{a_{11} a_{22}}} \ll 1$, and a variable $\xi$ being a scaled ratio of atom numbers (or atomic densities), $\xi =\frac{n_2 \sqrt{a_{22}}}{n_1 \sqrt{a_{11}}}$. After neglecting corrections of higher order in $\delta b$, Eq.(\ref{volume+}), can be brought to the form:
\begin{equation}
\label{magic_ratio}
 \frac{1}{2}\delta \xi^2 <  \delta  b
\end{equation}
where  $\delta \xi = \xi -1$. Obviously $\delta \xi$ is the second small parameter of the theory.

In view of Eq.(\ref{magic_ratio}) it is reasonable to assume that at equilibrium $\delta \xi \simeq 0$ so the ratio of atom numbers (and the ratio of equilibrium densities) is approximately equal to:   
\begin{equation}
\label{hard_mode}
\frac{N^0_1}{N^0_2}=\frac{n^0_1}{n^0_2}= \sqrt{\frac{a_{22}}{a_{11}}}=s
\end{equation}
This is a basic assumption of the analysis in \cite{Petrov15}.  Note that condition Eq.(\ref{hard_mode}) eliminates the hard mode contribution to the energy density Eq.(\ref{e-density}). Only soft mode and LHY energies remain. Using Eqs.(\ref{hard_mode}) and (\ref{volume}) the equilibrium densities $n^0_i$ of a droplet can be well approximated by:
\begin{eqnarray}\label{n01}
n^0_1 a^3_{11}=\left(\frac{2}{3c}\right)^2  \frac{\delta b^2}{(1+s)^5}
\\
n^0_2 a^3_{22}=\left(\frac{2}{3c}\right)^2  \frac{\delta b^2}{(1+\frac{1}{s})^5}
\label{n02}
\end{eqnarray}\\
If $\delta \xi = 0$ then the ratio of densities of the components is equal to the `magic' value $s$ at which the hard mode contribution to the mean field energy vanishes.
Therefore $\delta \xi$ measures a deviation of a droplet's density from this ratio. On the other hand it is easy to check that this parameter equals to fluctuations of density of the hard mode. If we define relative fluctuations of the densities $\delta_i$  in each mode via the relation: $n_i=n^0_i(1+\delta_i)$, then $\delta \xi$ measures the difference between these fluctuations: 
\begin{equation}
\delta \xi = \delta_1 - \delta_2
\end{equation}

So far we have only made use of the condition that a stable droplet has a vanishing pressure $p=0$ which stabilizes its volume $V=V_u(N_1,N_2)$. The total energy of droplets, $E(N_1,N_2)=E_u(N_1,N_2,V_u(N_1,N_2))$, becomes a function of number of atoms only. We are thus on the same footing as in the situation of a droplet having a nonuniform density profile where there is no need to introduce a volume as a free parameter.

Let us observe that the two functions $E(N_1,N_2)$ and $E_u(N_1,N_2,V)$ are different because in the latter case $V$ is an independent variable as opposed to the previous case where volume is a well-defined function of $N_1,N_2$, Eq.(\ref{volume}). This leads to two different definitions of chemical potentials. One is the $\mu_i$ given by Eq.~(\ref{mui1}), $\mu_{i} = \partial E/\partial N_i$, and the second one is given by Eq.(\ref{mui2}), $\mu_{i,u} = \partial E_u/\partial N_i$.
The relation between these two is given by 
\begin{eqnarray}
\mu_i  &\equiv& \frac{\partial E(N_1,N_2)}{\partial N_i}=\frac{\partial E_u(N_1,N_2,V_u)}{\partial N_i}
-p_u \frac{\partial V_u}{\partial N_i}
\nonumber\\
&=&\frac{\partial E_u(N_1,N_2,V_u)}{\partial N_i} \equiv \mu_{i,u},
\end{eqnarray}
where $p_u=\frac{\partial E_u(N_1,N_2,V)}{\partial V}|_{V=V_u}$ is pressure of a droplet with volume $V_u(N_1,N_2)$. For a stable droplet this pressure vanishes $p_u = 0$, therefore there is no additional energy cost related to change of volume. Both definitions of chemical potential are equivalent, $\mu_i = \mu_{i,u}$.
The stability conditions Eqs.(\ref{stab_cond1}), (\ref{stab_cond2}) are valid also in the case when droplet densities are approximated by constant functions and volume is introduced as an additional free parameter. 

\begin{figure}[htb]
    \centering
    \includegraphics[width=0.40\textwidth]{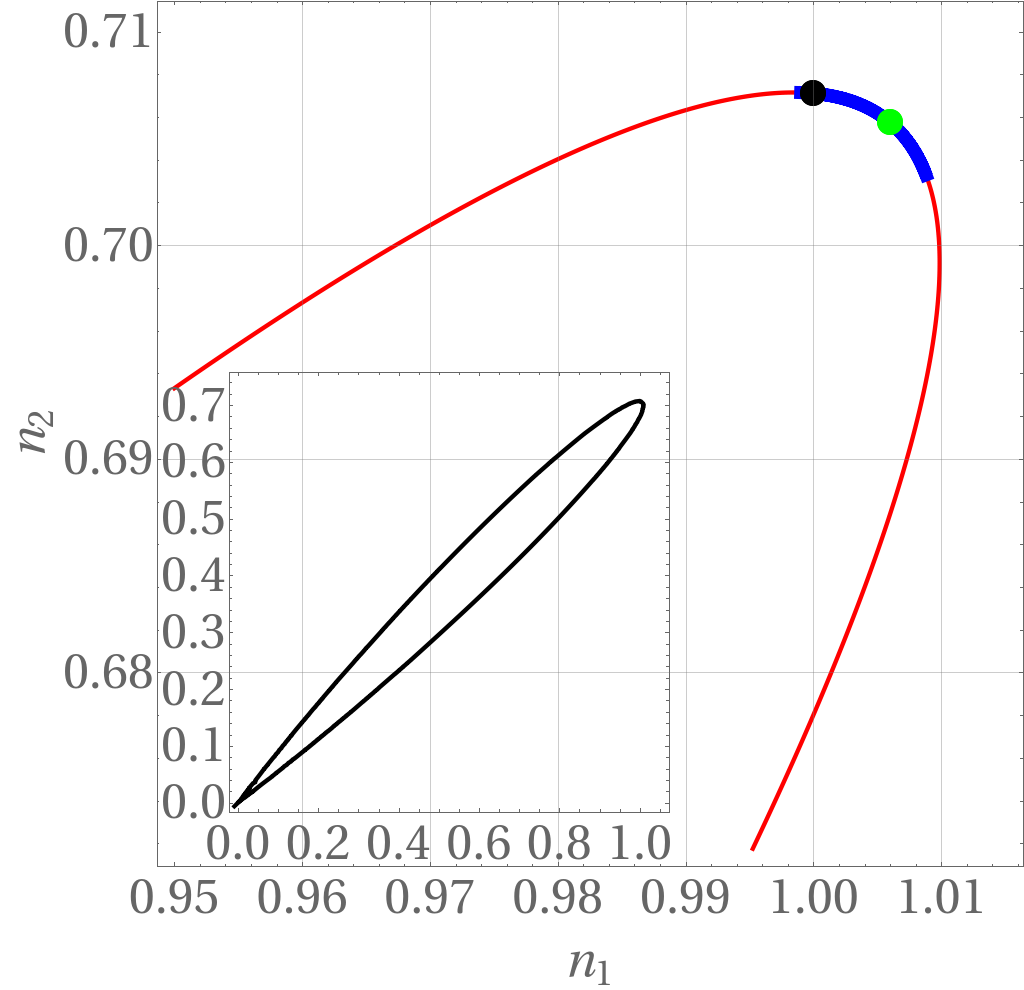}
    \includegraphics[width=0.41\textwidth]{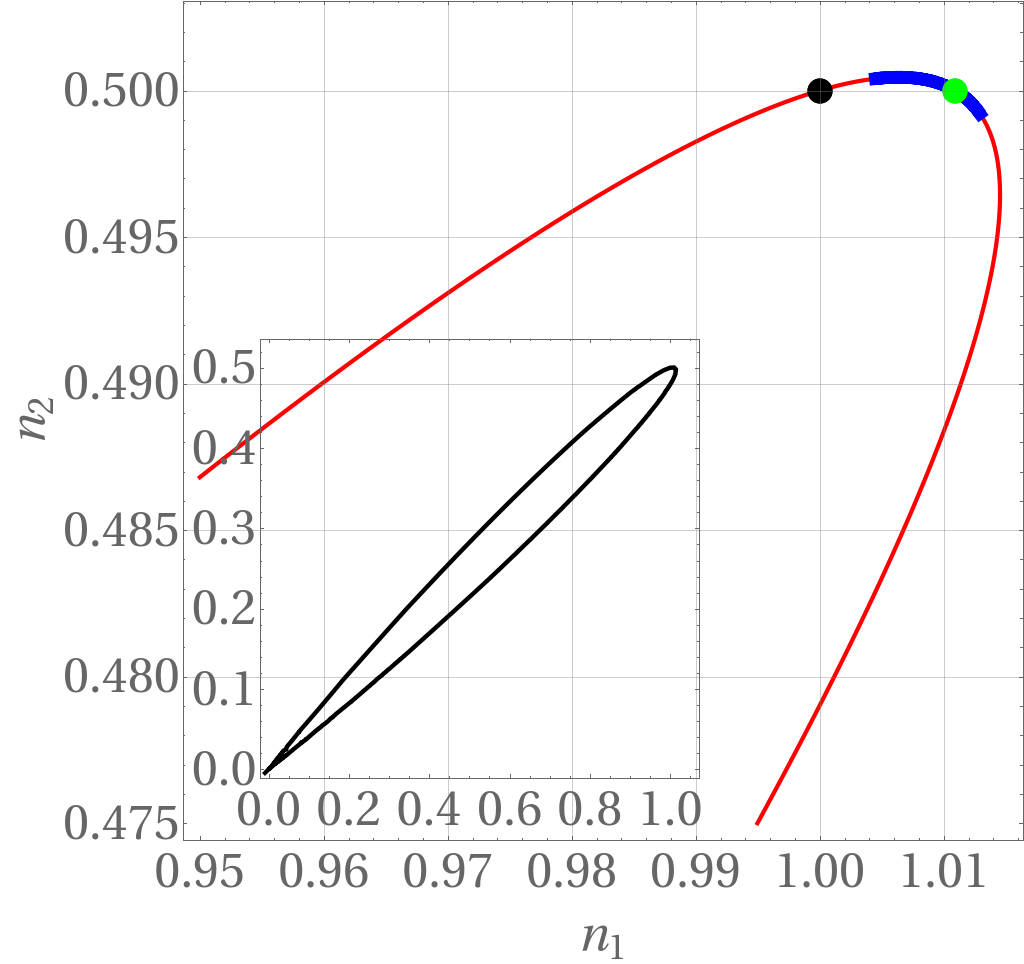}
    \caption{Solutions of Eq.~(\ref{pressure}) in the form of contour plots in the $n_1 - n_2$ plane. 
we show the tip of $p=0$ isobar where by blue color we indicate the stable region as given by $\mu_1<0$ and $\mu_2<0$ constraints, Eqs.(\ref{limits1},\ref{limits2}). In the inset we show the full zero pressure isobar which has a shape of the elongated loop. By the black dot we indicate the standard solution to the stability problem according to Eq.(\ref{hard_mode}). We consider two cases: (i) 
$s=\sqrt{2}$ (left panel). The standard solution  is located at the border, but inside the stable region. The solution given by Eq.(\ref{debraj_cond}) marked in green is at the centre;  (ii) 
$s=2$ (right panel). In this case the standard solution is located outside the stable region.   The solution given by Eq.(\ref{debraj_cond}) remains well within the limit of stability. Densities are in units of $n_{10}=\frac{25\pi}{1024}\frac{\delta a^2}{a_{11}^5s^2(1+s)^5}$}
    \label{fig:cisnienie}
\end{figure} 

Utilizing the explicit form of the energy density functional Eq.(\ref{e-density}), stability conditions of a quantum Bose-Bose droplet can be written as follows:\\
(i) the pressure vanishes , $p(n_1,n_2)=0$, Eq.(\ref{pressure}):
\begin{eqnarray}
\frac{p}{e_0} &=&- \delta b + \frac{1}{2} \frac{\delta \xi^2}{\xi}  +
\eta \left(   \frac{1}{\sqrt{\xi s}} + \sqrt{\xi s }   \right)^{5/2}= 0
\end{eqnarray}
(ii) both chemical potentials are negative:
\small
\begin{equation} 
\label{limits1}
\frac{\mu_1n_1}{e_0} =  -  \delta b -
\frac{\delta \xi}{\xi} 
+\frac{5}{3} \eta \left( \sqrt{\xi s} + \frac{1}{\sqrt{\xi s}} \right)^{3/2}  \frac{1}{\sqrt{\xi s}} <  0
\end{equation}
\begin{equation}
\label{limits2}
\frac{\mu_2n_2 }{e_0} =-  \delta b + \delta \xi +
\frac{5}{3} \eta \left( \sqrt{\xi s} + \frac{1}{\sqrt{\xi s}} \right)^{3/2} \sqrt{\xi s} < 0
\end{equation}
\normalsize
where we introduced: $e_0 = \frac{4 \pi \hbar^2}{m} \sqrt{a_{11}a_{22}} n_1 n_2$, and  $\eta=\frac{3c}{2} \left( n_1a^3_{11}n_2 a^3_{22}\right)^{1/4}$. By expanding the above equations to leading order in the small parameter $\delta \xi=\xi -1 $ we get the equation corresponding to the $p=0$ isobar in the $n_1-n_2$ plane:
\begin{equation} \label{rr1}
\eta =\frac{3c}{2} \left( n_1a^3_{11}n_2 a^3_{22}\right)^{1/4} = \frac{\delta b - \frac{1}{2} \delta \xi^2 }{  \left(   \frac{1}{\sqrt{s}} + \sqrt{s }   \right)^{5/2}  }
\end{equation}
Similar expansion allows for approximate but analytic determination of conditions limiting the region of stability of a quantum droplet with respect to evaporation, Eqs. (\ref{limits1},\ref{limits2}). The region of corresponding parameters forms a segment of $p=0$ isobar where the ratio of densities are limited as follows:
\begin{equation}
\label{stability}
- 1  + \frac{5}{3} \frac{s}{1 + s } < -\frac{\delta \xi}{\delta b} <  1  - \frac{5}{3}  \frac{1 }{1 + s  } 
\end{equation}
Equations (\ref{rr1}),(\ref{stability}) summarize the main results of our paper. They specify the region in the density-density plane where stable droplets exist. These results should be compared to other approximate expressions existing in the literature.

\subsection{Comparison with previous results} 

In this subsection we compare our results with previous formulae existing in the literature. Historically the first expression for droplets density was presented in \cite{Petrov15}. Based on: (i) assumption that hard mode contribution to the system energy vanishes at equilibrium, and (ii) condition of vanishing pressure, the equilibrium densities are found - see Egs.(\ref{n01}),(\ref{n02}). The solution gives $n_1=n^0_1$, $n_2=n^0_2$. Because all approaches discussed below use condition of vanishing pressure therefore a value of the parameter $\delta \xi$ is a perfect measure of differences between various results. Approximations of \cite{Petrov15} give: 
\begin{equation}
\label{petrov}
\delta \xi = \left(\frac{n_1}{n^0_1}-\frac{n_2}{n^0_2}\right) = 0
\end{equation}
This is the simplest but very good estimation of ratio of densities of large droplets confirmed in experiments \cite{Tarruell18,Cabrera18,Fattori18}.

Another solution is given in \cite{Rakshit19,Rakshit19low}.  Although results obtained there are dedicated to a Bose-Fermi mixture, but by following the main lines of the approach one can easily adapt the solutions of \cite{Rakshit19,Rakshit19low} to a Bose-Bose system. The energy functional of a Bose-Fermi mixture is not a quadratic form of densities and splitting of energy into hard and soft modes is not possible. Thus the approach of Ref.\cite{Petrov15}  cannot be used there. Instead, the results are obtained in the limit of infinite uniform system.  The issue of a finite volume does not have to be addressed at all then. In such a situation only intensive quantities make sense. These are the energy density and pressure. 

The stability problem is defined as a problem of finding a constrained minimum of the energy density under condition of vanishing pressure. This condition ensures that there are no net internal forces acting on a fictitious surface inside the bulk of a droplet. This is the same condition which fixes the volume of a finite homogeneous droplet, Eq.(\ref{pressure}). Additionally, at a minimum of energy density $\varepsilon(n_1,n_2)$ any infinitesimally small variation of atomic densities cannot change the energy:
\begin{equation}
d\varepsilon = \mu_1 dn_1+\mu_2 dn_2 =0
\label{min_e}
\end{equation}
To stay on the $p(n_1,n_2)=0$ isobar the variations of both densities $dn_1$ and $d n_2$ must be related:
\begin{equation}
dp=\frac{\partial p}{\partial n_1} dn_1+\frac{\partial p}{\partial n_2}dn_2=0.
\label{countour}
\end{equation}
Combining condition Eq.(\ref{countour}) with  Eq.(\ref{min_e}) the following equation for the constrained minimum of energy density, $\epsilon(n_1,n_2)$, can be found:
\begin{equation}
\label{debraj_cond}
\mu_1 \frac{\partial p}{\partial n_2} - \mu_2 \frac{\partial p}{\partial n_1}=0.
\end{equation}
When accompanied by $p(n_1,n_2)=0$ equation the densities of both species can be found.

Eq.(\ref{debraj_cond}) accompanied by $p(n_1,n_2)=0$ condition can be applied to the Bose-Bose mixtures. Again, expressing derivatives of pressure contributing to the above equation in terms of $\xi$ and expanding Eq.(\ref{debraj_cond}) in the small parameter $\delta \xi$  the approximate solution for densities of a stationary droplet is:
\begin{eqnarray}
\label{debraj}
\delta \xi = \left(\frac{n_1}{n^0_1}-\frac{n_2}{n^0_2}\right)=\delta b \left( \frac{1-s}{1+s} \right).
\end{eqnarray}
This solution meets the stability criteria defined here, Eq.(\ref{stability}). Independently of the value of the parameter $s$,  Eq.(\ref{debraj}) predicts droplet densities very close to the center of the stability region. This solution is marked by a green dot in Fig.~(\ref{fig:cisnienie}).


Evidently both formulae Eq.~(\ref{petrov}) and Eq.(\ref{debraj}) are equivalent if intraspecies interactions are equal, i.e. if  $s=1$. Note that Eq.(\ref{debraj}) confirms small contribution of hard mode excitations to the densities of a stable droplet. 

The situation is different for $s>3/2$. Then the standard result $\delta \xi = 0$ given by Eq. (\ref{petrov}) is outside the stability region given by Eq.~(\ref{stability}). Thus, for sufficiently strong asymmetric intraspecies interaction the standard solution does not support a stable droplet.

\begin{figure*}[htb]
    \centering
    \includegraphics[width=0.40\textwidth]{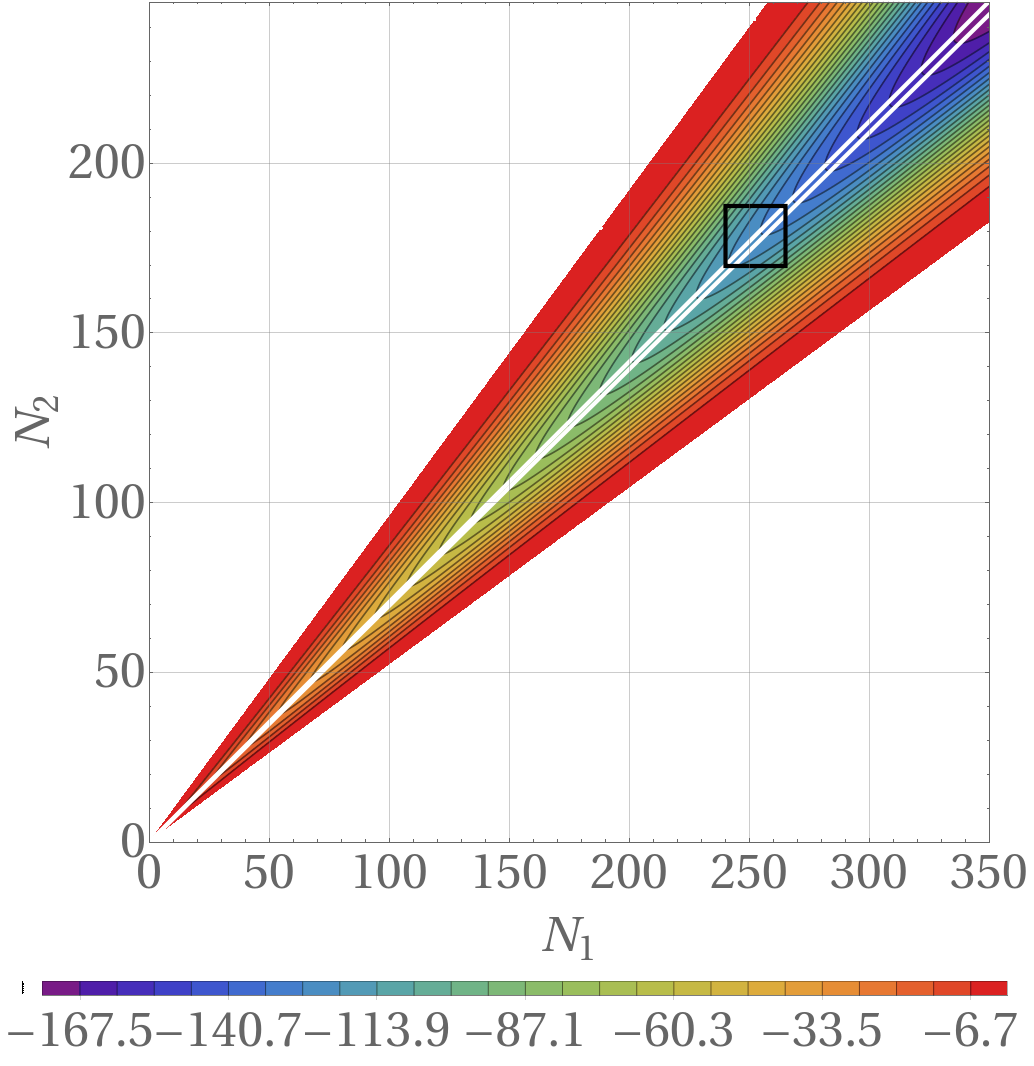}
    \includegraphics[width=0.40\textwidth]{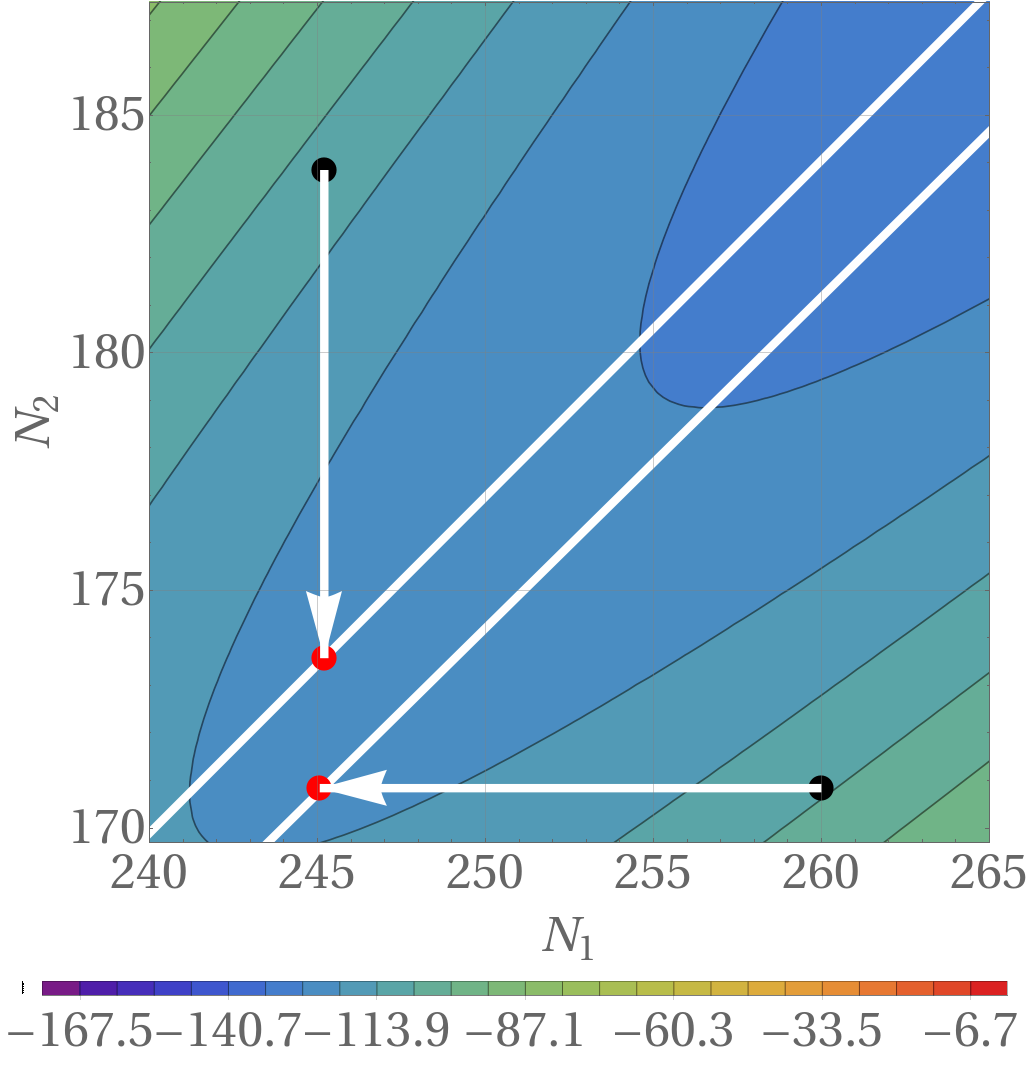}
    \caption{The total energy as a function of the number of particles in every component for unequal
    intraspecies interaction, $s=\sqrt{2}$. Left panel: Coloured region corresponds to such a composition of the mixture for which  $p=0$ condition can be met. The isobar $p=0$ shown 
    in Fig.(\ref{fig:cisnienie}) becomes here the interior of the angular region given by 
    Eq.(\ref{volume+}). White lines indicate the edges of the zone of stable droplets where 
    $\mu_1<0,\mu_2<0$. The rectangle at the center indicates the region which we zoom-in in the right panel. Right panel: Zoom of the energy landscape in $N_1-N_2$ plane. It illustrates adiabatic evolution of two initial states $(N_1^{ini},N_2^{ini})$ marked by black dots. Evolution towards the state of minimal possible energy constrained by initial atom numbers cannot have any positive-valued gradient component of the chemical potential vector $(\mu_1,\mu_2)$. The white arrows show trajectories towards the final state $(N_1^{fin},N_2^{fin})$ (red dots) of lowest possible energy for the assumed arrangement. Please note that only the edges of the stability region can be reached. Getting into the interior of this region requires increasing the number of atoms of at least one kind. On the contrary, all systems having initially a number of particles corresponding to the area between the white lines are stable against small perturbations. The number of atoms is expressed in convenient units of Ref. \cite{Petrov15}. Therefore, 'the real' number of atoms is equal to $N_r=N\cdot n_{10}\tilde{r}^3\approx N\cdot6300$, where $\tilde{r}=\sqrt{\frac{3}{2}\frac{s+1}{4\pi|\delta a|n_{10}}}$ is the length unit}
    \label{fig:energia}
\end{figure*} 

The results are illustrated in Fig.(\ref{fig:cisnienie}) where we show the stability diagram in a plane of atomic densities, $n_1$ and $n_2$. For comparison we present the two cases: $s=\sqrt{2}$ and $s=2$. The $p=0$ isobar has the form of a closed loop originating at the center of the coordinate system -- see inset in Fig.~(\ref{fig:cisnienie}). The region which is stable with respect to atom losses, ($\mu_1,\mu_2\leq0$), Eqs.~(\ref{limits1},\ref{limits2}), is located close to the tip of the loop which we zoom-in in the main frame. This is the part of the isobar marked in blue.  By green dot we mark the solution corresponding to the global minimum of an infinite system as suggested in \cite{Rakshit19, Rakshit19low} and given by Eq.~(\ref{debraj}). This result is well in the stable part of the diagram regardless the interactions. The standard solution of \cite{Petrov15}, Eq.(\ref{hard_mode}), is indicated by a black dot. We stress that when the disproportion of intraspecies interactions is too large ($s=2$) the standard solution of Ref.\cite{Petrov15} is out of the stability region. 

\subsection{Stable equilibrium for given initial particle number}
In the last part of the paper we go back to the problem which was the inspiration for our study. We address the question asked at the beginning of this work, i.e. we are going to show what is a minimal energy state which can be reached having at disposal $N_1^{ini}$ atoms of the first kind and $N_2^{ini}$ atoms of the second kind allowing for throwing away some of them. 
			
The solution to this problem is illustrated in Fig.~(\ref{fig:energia}) which shows the total energy of the system $E(N_1,N_2,V(N_1,N_2)) = V(N_1,N_2) \varepsilon(N_1/V,N_2/V)$ in the plane of extensive quantities $N_1,N_2$. If one has initially a two component mixture with $(N_1^{ini},N_2^{ini})$ atoms then the droplet formed would be in general a mixture of $(N_1^{fin},N_2^{fin})$ atoms of both kinds. To find the droplet of the lowest energy among all possible final states of droplets composed with number of atoms limited by the initial values, $N_1^{fin} \leq N_1^{ini}$ and  $N_2^{fin} \leq N_2^{ini}$ we directly examine the region of energies in the relevant rectangular domain in $N_1-N_2$ plane:
\begin{eqnarray}
0 \leq N_1 \leq N_1^{ini},\\
0 \leq N_2 \leq N_2^{ini}.
\end{eqnarray}

In Fig.~(\ref{fig:energia}) the initial composition of droplet is marked by a black dot.  White vertical and horizontal arrows point to the final states $(N_1^{fin},N_2^{fin})$ which minimize the energy constrained according to the previous discussion. We consider two situations. The first one is that $N_2^{ini}/N_1^{ini}$ is so large that $- \delta\xi/\delta b > 1-5/3(1/(1+s))$, i.e. the second component of the mixture is a strongly excessive one. In such a case the excessive atoms simply evaporate until the system reaches the boundary of the stable region, vertical arrow in the figure. It is worth mentioning that the number of minority atoms is conserved. Further evaporation stops when the border of the stability sector is reached. This is because equilibrium results from a competition of the two tendencies: the system tends to decrease the chemical potential (the energy per particle) as much as possible and simultaneously to keep as many atoms as possible, since total energy is extensive. 

The second case shown in Fig.~(\ref{fig:energia}) relates to the situation where the first component dominates, i.e. if $- \delta\xi/\delta b < -1+5/3(s/(1+s))$. The scenario described above repeats. Excessive atoms of the first component evaporate, while the number of atoms in the second component remains constant (horizontal white arrow in figure). This process stops while reaching the border of the stable sector. 

If initially the system is prepared in the stable zone, i.e. in the area limited by the two white lines in Fig.(\ref{fig:energia}) it will not evaporate atoms at all. We have to add that the present discussion is based on stationary stability analysis and no time dynamics was considered at all. Therefore all our conclusions, in particular these invoking dynamic processes such as evaporation, implicitly assume that the system remains at equilibrium and adiabatically follows the state determined by external parameters and temporal number of atoms. For the same reason we are not able to discuss a situation when the initial number of atoms is outside of the coloured angular sector in Fig.(\ref{fig:energia}) indicating $p=0$ zone. The outer part is a totally unstable sector and releasing atoms from the trap while parameters are in this range will trigger a violent dynamics. Only dynamical studies of the process of droplet formation might give the state of an eventually formed droplet. 

\section{Conclusions}
In this paper we specified stability conditions for a self-bound two-component droplet. The conditions are not related to any particular form of the energy density functional, and can be applied to other systems like for instance  Bose-Fermi mixtures. The case of Bose-Bose droplets was studied in detail. In contrast to the standard solution of  Ref.\cite{Petrov15} we show that stable droplets' densities, for fixed values of interactions strength, can take values from some finite range of parameters, thus there is no unique droplet solution. This regime of allowed densities is however rather small and deviations from the standard solution are limited particularly for similar strengths of intraspecies interactions. In the limit of large droplets, when kinetic energy can be neglected, we found a very useful analytic expression for the boundaries of the stability zone. We have shown that if intraspecies interactions are very different from each other then the prediction of Ref.\cite{Petrov15} is outside the stability sector.

\begin{acknowledgments}
We thank Filip Gampel for discussion and proofreading of the manuscript.
Authors acknowledge support from National Science Center (Polish)) grant No. 2017/25/B/ST2/01943. 
\end{acknowledgments}

\bibliographystyle{apsrev4-1}

\bibliography{main}






\end{document}